\documentclass{aa}
\usepackage{graphicx}
\usepackage{lscape}

\begin{document}

\def\kms{km~s$^{-1}$}
\def\ctq{CTQ~247}
\def\q{Q1331+17}
\def\cm2{cm$^{-2}$}
\def\hkpc{$~h_{50}^{-1}$ kpc}
\def\icm{cm$^{-2}$}
\def\lya{Ly$\alpha$}
\def\lyb{Ly$\beta$}
\def\lyc{Ly$\gamma$}
\def\nfive{\ion{N}{v}}  
\def\feone{\ion{Fe}{ii}{ $\lambda2344$}}  
\def\fetwo{\ion{Fe}{ii}}
\def\fethree{\ion{Fe}{iii}}  
\def\fefour{\ion{Fe}{ii}{ $\lambda2586$}}  
\def\fefive{\ion{Fe}{ii}{ $\lambda2600$}}  
\def\fesix{\ion{Fe}{ii}{ $\lambda1608$}}  
\def\sione{\ion{Si}{iv}{ $\lambda1393$}}
\def\sitwo{\ion{Si}{ii}}
\def\sithree{\ion{Si}{iii}}
\def\sifour{\ion{Si}{iv}}
\def\mgone{\ion{Mg}{ii}{ $\lambda2796$}}  
\def\mgtwo{\ion{Mg}{ii}{ $\lambda2803$}}  
\def\mgthree{\ion{Mg}{i}{ $\lambda2852$}}  
\def\alone{\ion{Al}{ii}{ $\lambda1670$}}  
\def\oone{\ion{O}{vi}{ $\lambda1032$}}  
\def\otwo{\ion{O}{vi}{ $\lambda1037$}}  
\def\cfour{\ion{C}{iv}}  
\def\cthree{\ion{C}{iii}}  
\def\ctwo{\ion{C}{ii}}  
\def\si4{\ion{Si}{iv}}  
\def\s6{\ion{S}{vi}}  
\def\o6{\ion{O}{vi}}  
\def\oone{\ion{O}{i}}  
\def\n5{\ion{N}{v}}  
\def\hone{\ion{H}{i}}  
\def\htwo{\ion{H}{ii}}  
\def\altwo{\ion{Al}{ii}}  
\def\althree{\ion{Al}{iii}}  
\def\none{\ion{N}{i} }  
\def\nhi{$N_{\rm \ion{H}{i}}$}
\def\c43{\ion{C}{iv}/\ion{C}{iii}}
\def\no{$\cdot\cdot\cdot$}
\def\ptwo{\ion{P}{ii}}  
\def\pthree{\ion{P}{iii}}  
\def\fetwo{\ion{Fe}{ii}}  
\def\nitwo{\ion{Ni}{ii}}  
\def\zntwo{\ion{Zn}{ii}}  
\def\mntwo{\ion{Mn}{ii}}  
\def\crtwo{\ion{Cr}{ii}}  
\def\stwo{\ion{S}{ii}}  
\def\sthree{\ion{S}{iii}}  
\def\arone{\ion{Ar}{i}}  
\def\artwo{\ion{Ar}{ii}}

\title{Distinct Abundance Patterns in Multiple Damped Ly$\alpha$
Galaxies: Evidence for Truncated Star Formation?\thanks{The
work presented here is based on data obtained at the ESO Very Large
Telescopes, programs  68.A-0361(A) and  66.A-0660(A).}  
}

\titlerunning{Abundances in Multiple DLAs}

   \author{Sebastian Lopez\inst{1}
          \and
          Sara L. Ellison\inst{2,3}
          }

   \offprints{S. Lopez}

  \institute{
	Departamento de Astronom\'{\i}a, Universidad de Chile, Casilla 36-D,
        Santiago, Chile \\
	\email{slopez@das.uchile.cl}
        \and
	European Southern Observatory, Casilla 19001, Santiago 19, 
        Chile
	\and
	Pontificia Universidad Cat\'olica de Chile, 
	Casilla 306, Santiago 22, Chile\\
        \email{sellison@astro.puc.cl}}

   \date{Received / Accepted}

\abstract{ Following our previous work on metal abundances of a double
damped \lya\ system with a line-of-sight separation $\sim 2\,000$ \kms
(Ellison \& Lopez 2001), we present VLT UVES abundances of 3 new
systems spanning a total of $\sim 6\,000$ \kms\ at $z \sim 2.5$ toward the
southern QSO CTQ247. These abundances are supplemented with echelle
observations of another `double' damped \lya\ system in the
literature. We propose a definition in terms of velocity shift of the
sub-class 'multiple damped \lya\ system', which is motivated by its possible
connection with large-scale structure. 
We find that the abundance ratio [S/Fe] is systematically
low in multiple systems compared with single systems, and
with a small scatter.  The same behavior is found in 2 more single DLA
systems taken from the literature that show evidence of belonging to a
galaxy group. Although [Si/Fe] ratios are also generally lower in
multiple DLAs than in single DLAs, the effect is less striking since the
scatter is larger and there are a number of low [Si/Fe] DLAs in the
literature.  We suggest that this can be explained with a combination
of detection bias and, to a lesser extent, the scatter in ionization
corrections for different absorbers.  We investigate whether
consistently low $\alpha$/Fe ratios could be due to dust depletion or
ionization corrections and find that the former effect would emphasize
the observed trend of low $\alpha$/Fe in multiple systems even
further.  Ionization may have a minor effect in some cases, but at a
level that would not change our conclusions.  We thus conclude that
the low $\alpha$/Fe ratios in multiple DLAs have a nucleosynthetic origin and
suggest that they could be explained by reduced star formation in
multiple damped \lya\ systems, possibly due to environmental effects.
There seems to be independent evidence for this scenario from the mild
odd-even effect and from the relatively high N/$\alpha$ ratios we
observe in these multiple systems.
}

\maketitle

\section{Introduction}

\begin{table*}
\begin{center}
      \caption[]{VLT Observations of \ctq}
         \label{tbl-1}
      \[
         \begin{tabular}{lccr}
            \hline
            \noalign{\smallskip}
{\rm Mode}&{\rm Wavelength}&{\rm Exp. Time}&{\rm Observing Date}\\
              &[nm]           &  [sec]        &                    \\
            \noalign{\smallskip}
            \hline
            \noalign{\smallskip}
FORS2 (GRISM 600B)  &345-590&600&Dec. 18 2000\\
FORS2 (GRISM 600R)  &525-745&400&Dec. 18 2000\\
UVES Dichroic (390+580)  & 328-445,476-684 &13\,500&Oct. 10, 11 and Nov.17, 2001\\ 
UVES Dichroic (437+860)  &376-500,667-1040&10\,800&Nov. 17 2001\\
            \noalign{\smallskip}
            \hline
         \end{tabular}
      \]
\end{center}
\end{table*}

Damped Lyman Alpha systems (DLAs) in the spectra of high-redshift
quasars have promised to be a powerful probe of early metal enrichment
in young galaxies. However, despite the considerable effort invested
to develop this technique, its full potential has yet to be realized.
Although  extensive observing campaigns at intermediate resolution have swollen
the list of known DLAs (\cite{corals1}, \cite{Lopez01} and
\cite{Peroux} are some of the most recent DLA surveys),
only a  fraction have been followed-up at high resolution
to study their chemical properties, investigate dust
content, photoionization and metallicity evolution. 
With the present sample of some $60$ abundance measurements some
broad trends have been identified (e.g., \cite{ucsd2}), but the 
global picture of star formation (SF) history deduced from these 
patterns of elemental abundances --likely affected by dust depletion
patterns (\cite{Hou})-- is still unclear.  In particular,
extracting coherent scenarios of SF histories in individual DLAs, or subsets of
absorbers remains a challenging prospect with current data.

One of the great advantages of studying high redshift galaxies
in \textit{absorption} is the lack of selection bias associated 
with cosmological
dimming: galaxies are selected based only on gas cross section,
irrespective of their intrinsic luminosity.  However, this feature
may turn out to be a double-edged sword in that extracting coherent
chemical enrichment patterns is almost certainly hindered by the mixed
morphological representation.  Undoubtedly, some of the scatter 
in relative abundances comes from variable dust depletion, although 
system-to-system differences persist even after correcting for 
grain fraction (Vladilo 2002).  In order to compare elemental trends
in DLAs in a similar manner to, for example, local stellar abundance
studies, a significant refinement in our approach is required. 
Such a refinement might include isolating particular subsets of
DLAs based either on their morphology or some other observed
property (e.g. spin temperature;  \cite{Chengalur}).

In this paper, we investigate whether a distinct enrichment trend
exists for absorption line systems with close companions in velocity
space, which we  
will refer to as `Multiple Damped Lyman Alpha systems', or MDLAs.

\section{The Sub-class 'MDLA'}
\label{def}

Lopez et al. (2001) identified the existence of three damped \lya\ absorbers
spanning a total of $v= 5\,900$~\kms\ at $z\sim 2.5$ toward the
southern quasar \ctq, the first ever example of a triple DLA; Ellison
et al. (2001b) discovered a double absorption systems for which 
$v= 1\,800$~\kms\ at
$z\sim 2$ toward Q2314$-$409; a similar velocity span as in \ctq\ is
covered by the two DLAs toward Q2359$-$02 at $z\sim2$
(\cite{Wolfe86}). Statistically, these are extremely improbable events
if the proximity of the systems is purely by chance, and only these
few cases are reported in the literature.  
It is therefore feasible
to postulate that in these cases the multiple absorption systems
may be in some way related, rather than chance alignments.
However, such large velocities are difficult to reconcile with today's galaxy
groups and  clusters --i.e., virialized entities-- 
or even with superclusters.
However, at $z>0$ current work on small angular fields shows evidence for 
very large structures on
comoving scales as large as $d \sim 100~h^{-1}_{100}$ Mpc 
($\Delta v\sim
10\,000$~\kms\ if purely due to Hubble flow) at various redshifts.
Evidence for large structure on many Mpc scales 
has been found in absorbing gas both
across (e.g.,~\cite{Williger02}; $d \sim 100~h^{-1}$ Mpc at $z\sim 1.3$ 
) and along the line of sight  
(\cite{Quashnock96}; $d \sim 100~h^{-1}$ Mpc at $z\sim 3$), in
super-clusters ($d\sim 20~h^{-1}$ Mpc at $z\sim 0.8$; \cite{hcc03}) and  
in Lyman-break galaxies (Steidel et al. 1998; $d \sim 10~h^{-1}$ Mpc
at $z\sim 3$ ). Large structure has also been reproduced 
in CDM N-body simulations ($d\sim 20~h^{-1}$ Mpc at $z\sim 1$;
\cite{Evrard02}).  It therefore seems a feasible 
(but not exclusive) possibility that absorption systems separated by 
several thousand \kms\ may be associated with very large scale structure.

Motivated both by the uniqueness of MDLAs and the possibility of 
large-structure, we
have embarked on a program to measure abundances in these
systems.  We will
define an MDLA as 2 or more DLAs spanning a velocity range

\begin{equation}
500<\Delta v<10\,000~{\rm km~s}^{-1}.
\end{equation}

While the upper limit is set to match the limits of the largest structures
known at high redshift, the lower velocity roughly corresponds to the 
largest asymptotic velocities in massive spirals (e.g. 
\cite{rig02})\footnote{This
lower limit will probably exclude the majority of bound galaxy 
satellites since most local
spirals' satellites have velocity differences of less than 300 \kms, 
\cite{zar97}}.  Note that this a
{\it working} definition only; although it might be possible that
MDLAs probe some kind of large structure (but see
\S~\ref{conclusions}), our definition does not necessarily imply that
they host the strong \ion{C}{iv} absorption used by Quashnock, vanden
Berk \& York (1996) 
or they are associated with the LBGs identified by \cite{ste96}.

From an instrumental point of view, identifying MDLAs in \lya\
--in contrast to using metal lines-- will certainly limit the
identification of systems separated by  $\Delta v \la 1\,000$ \kms.  
If the discovery spectrum has poor S/N, obtaining line parameters
for close MDLAs will require 
both the detection of metal lines, and a good continuum
estimation\footnote{Two $N($\ion{H}{i}$)=2\times10^{20}$ \icm\ DLAs
  separated by 
$\Delta v =1\,000$ \kms\ will mimic a single system in FWHM $\sim5$ \AA\
data if S/N $\la 20$, but for $\Delta v \sim 500$ \kms\ they will do
so unless S/N $\ga 50$.}.  

For $N($\ion{H}{i}$)$ criterion, we will relax the canonical
limit to include systems with $N($\ion{H}{i}$) >10^{20}$ \icm. Not
only do systems down to this limit continue to exhibit clear damping
wings, but in the
following we show that in such range ionization does not yet play a
significant role when deriving abundances from low-ions.  

The first in-depth study of an MDLA was presented by Ellison \& Lopez
(2001; hereafter Paper I).  There, the two DLAs toward Q2314$-$409
were noted to have low [S/Fe] ratios, a  result which we suggested was
due to 
the possible impact of environment on the chemical abundances in DLA
protogalaxies. In the present work, we extend the study of abundances
in MDLAs by 
investigating the triple DLA toward CTQ247 with new echelle data, and
including literature abundances of the double system toward
Q2359$-$02. These new abundances reinforce 
our previous suggestion of peculiar relative
abundances in MDLAs.  

After presenting the new data in
\S~\ref{data}, the discussion of the abundances is given in
\S~\ref{abundances}, where we attempt a statistical comparison
between the 2 populations, and assess possible systematic effects that
may bias our result. A discussion on the 
nature of MDLAs is outlined in \S~\ref{conclusions}

\section{Data Analysis}
\label{data}
\subsection{Observations and Data Reduction}


\begin{figure*}
\centerline{\rotatebox{0}{\resizebox{15.0cm}{!}
{\includegraphics{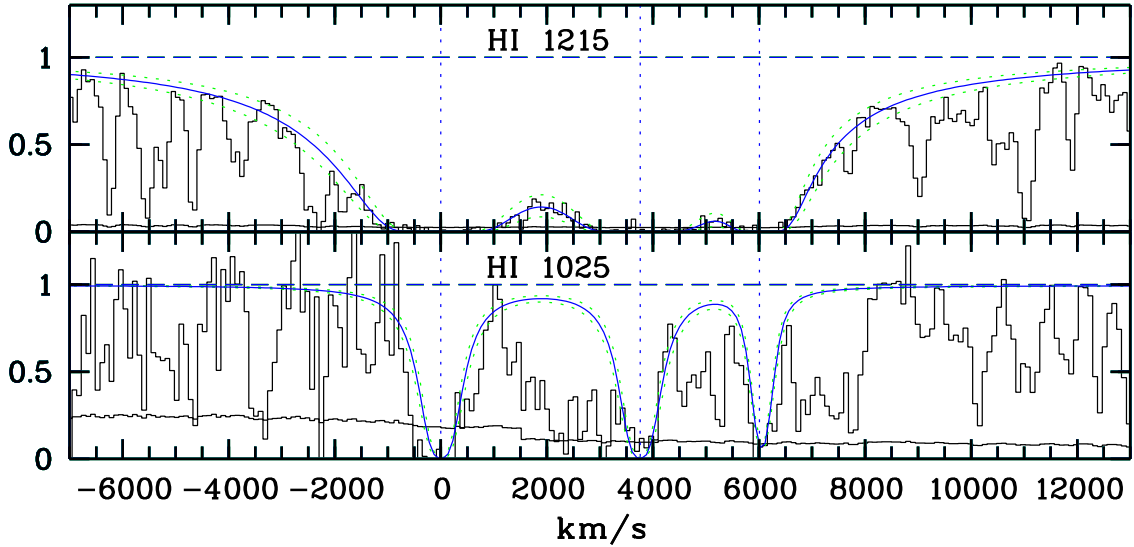} }}}
\caption{\label{fig_HI} Normalized flux and $1 \sigma$ VLT FORS2
spectrum of CTQ 247 at FWHM $=3$   
\AA\ resolution showing
  the \lya\ and \lyb\ lines of the triple DLA. Overlaid is
  a synthetic Voigt profile of three systems with
  $N($\ion{H}{1})$=10^{21.13}$, 
  $10^{21.09}$, and $10^{20.47}$ \icm, centered at $z_{abs}=2.5505$, $2.5950$,
  and $2.6215$. The dotted lines are $\pm 0.1$ dex deviations for all
$N($\ion{H}{i}) values.}   
\end{figure*}

\ctq\ was observed in service mode with the UVES instrument at the ESO
Kueyen telescope in October and November 2001 under good seeing
conditions ($0.6\arcsec-1.0\arcsec$).  With the 390+580 and 437+860
dichroic modes we covered from 328 to 1020 nm with two gaps at 576-583
nm and 852-866 nm.  The exposure times were $13\,500$ seconds for
dichroic 1 and $10\,800$ seconds for dichroic 2 (Table~\ref{tbl-1}).

After the usual fashion of bias-subtracting and flat-fielding of the
individual CCD frames, the echelle orders were extracted and reduced
interactively with the UVES pipeline routines (\cite{bal00}).  The reference
Th-Ar spectra used for wavelength calibration were taken after each
science exposure.  The wavelength values were converted to
vacuum heliocentric values and each order of a given instrumental
configuration was binned onto a common linear wavelength scale of
$0.04$ \AA~pixel$^{-1}$. The reduced orders were then added with a
weight according to the inverse of the flux variances. Finally, the
flux values were normalized by a continuum that was defined using
cubic splines over featureless spectral regions. The spectral
resolution is FWHM $\sim 6.7$ \kms, while the typical signal-to-noise ratio
per pixel is S/N$\sim 35-40$. 

In addition to the UVES data, a lower resolution (FWHM $\approx 3$ \AA)
spectrum of \ctq\ was obtained on December 18 2000 using FORS2 at the
ESO Kueyen Telescope. This spectrum was used to better define  the
quasar continuum in the spectral region around the damped \lya\
lines.


\subsection{Column densities}

We used FITLYMAN and VPFIT\footnote{Available at http://www.ast.cam.ac.uk/\~{}rfc/vpfit.html } to fit the line profiles with theoretical
Voigt profiles.  All fits were unconstrained in redshift, Doppler
width and column density, unless otherwise stated.  In general, we
prefer this approach over the apparent optical depth method
(AODM; Savage \& Sembach 1991) when velocity components are not  
resolved as is sometimes the case.  In addition, when dealing with
transitions that lie in the Ly$\alpha$ forest, such as the \ion{S}{ii}
triplet, fitting provides important information to the extent of
possible blending.  Despite this, however, we did use the AODM 
for transitions where the fit failed due to line blending of many components 
or poor S/N.   
We adopted the up-to-date $f$-values listed
in~\cite{ucsd1} and the
solar abundances by \cite{gs98}, with updates for O and
N by \cite{hol01}.  Table~\ref{abund_tab_ctq} 
lists the column densities and derived abundances of all elements 
covered by our observations.

The
\ion{H}{i} column density was determined by fitting 3 components to the 
damped \lya\ and \lyb\ profiles in the FORS2 spectrum. The  choice of
the low resolution spectrum reduces the uncertainties inherent to
determining $N($\ion{H}{i}$)$ from echelle data, such as badly defined
continuum where the damping wings extend over more than one order. 
We obtained $N($\ion{H}{i}$)=10^{21.13}$,
$10^{21.09}$, and $10^{20.47}$ \icm, for the 3 systems centered at
$z_{abs}=2.5505$, $2.5950$ and $2.6215$. The theoretical profiles are
superposed on the data in Fig.~\ref{fig_HI}. The internal fit errors
were 0.02 dex for all 3 \ion{H}{i} measurements but we believe a more
realistic error is given by $\pm 0.1$, which is shown by the dotted
curve in the figure. Although the noisy data around \lyb\ 
constrains \ion{H}{i} only marginally, this error seems quite
safe, given the higher S/N at \lya.  

We next give a short description of the metal line fits in each of the 3
DLAs which we henceforth refer to as CTQ~247A,B and C.

\paragraph{CTQ247A ($z_{abs}=2.5505$):}


\begin{figure}
\centerline{\rotatebox{0}{\resizebox{8.0cm}{!}
{\includegraphics{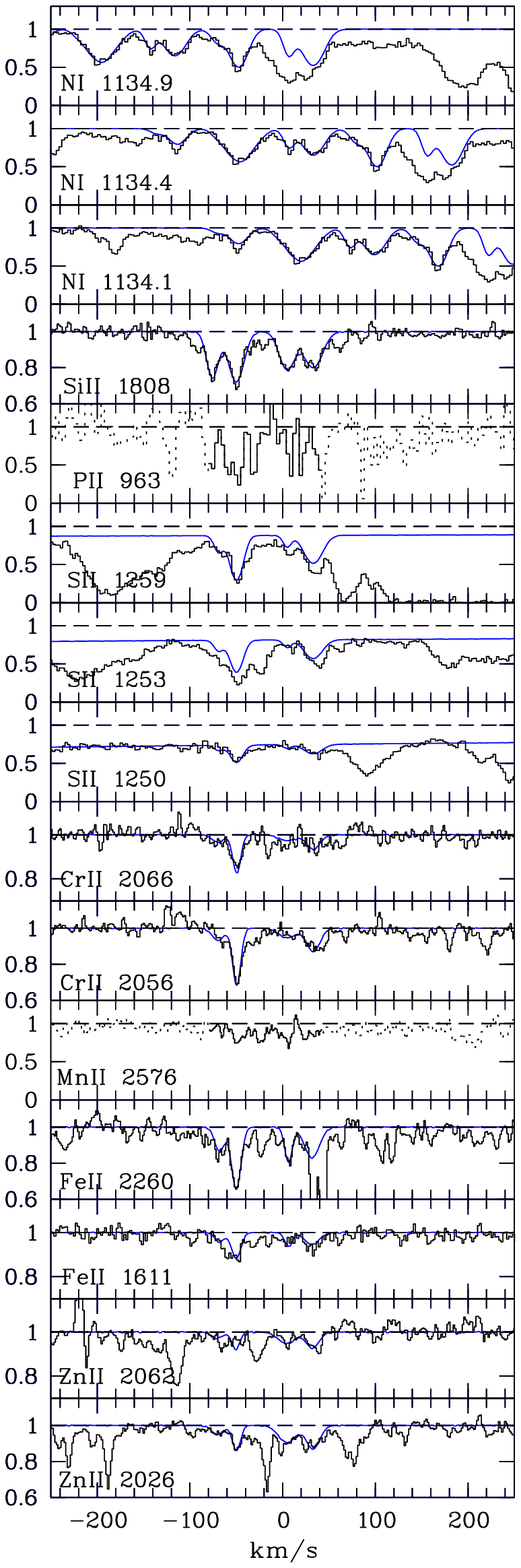} }}}
\caption{\label{fig_A}  UVES spectrum
  showing nonsaturated transitions in   
CTQ247A ($z_{abs}=2.5505$). The solid line in the \ion{P}{ii} and
\ion{Mn}{ii} panels  
indicates the velocity region used for the AODM.}
\end{figure}

For this DLA we have identified nonsaturated transitions of 
\ion{N}{i}, \ion{Si}{ii}, \ion{S}{ii}, \ion{Cr}{ii}, \ion{Fe}{ii}, and
\ion{Zn}{ii}, all of which show 4 velocity components.  
More (weaker) components are revealed by stronger transitions, e.g. 
\ion{Fe}{ii} $\lambda 2344$, but these do not contribute significantly to
the total column densities. Therefore, in order to avoid introducing
too many parameters at intrinsically weak transitions in relatively
noisy parts of the spectrum, we proceeded by fitting 4 components. 
Fig.~\ref{fig_A} shows the nonsaturated transitions used and their
fits. From an analysis of the weaker components we estimate that
the obtained column densities encompass $\sim 90$ \% of the total
column density; exclusion of weaker components will lead to only a
$\sim 0.05$ dex  
difference at $\log N=15$ \icm.   The
\ion{S}{ii} $1260$ triplet lines fall on the red wings of the damped
\lya\ lines; consequently, they were fitted simultaneously with the
\ion{H}{i} fixed at the  values found for the FORS data, hence the
sub-unity values of the continuum in Fig.~\ref{fig_A}. 
In addition, we determine an AODM measurement for \ion{Mn}{ii} and an
upper limit for \ion{P}{ii}.

\paragraph{CTQ247B ($z_{abs}=2.5950$):}

\begin{figure}
\centerline{\rotatebox{0}{\resizebox{8.0cm}{!}
{\includegraphics{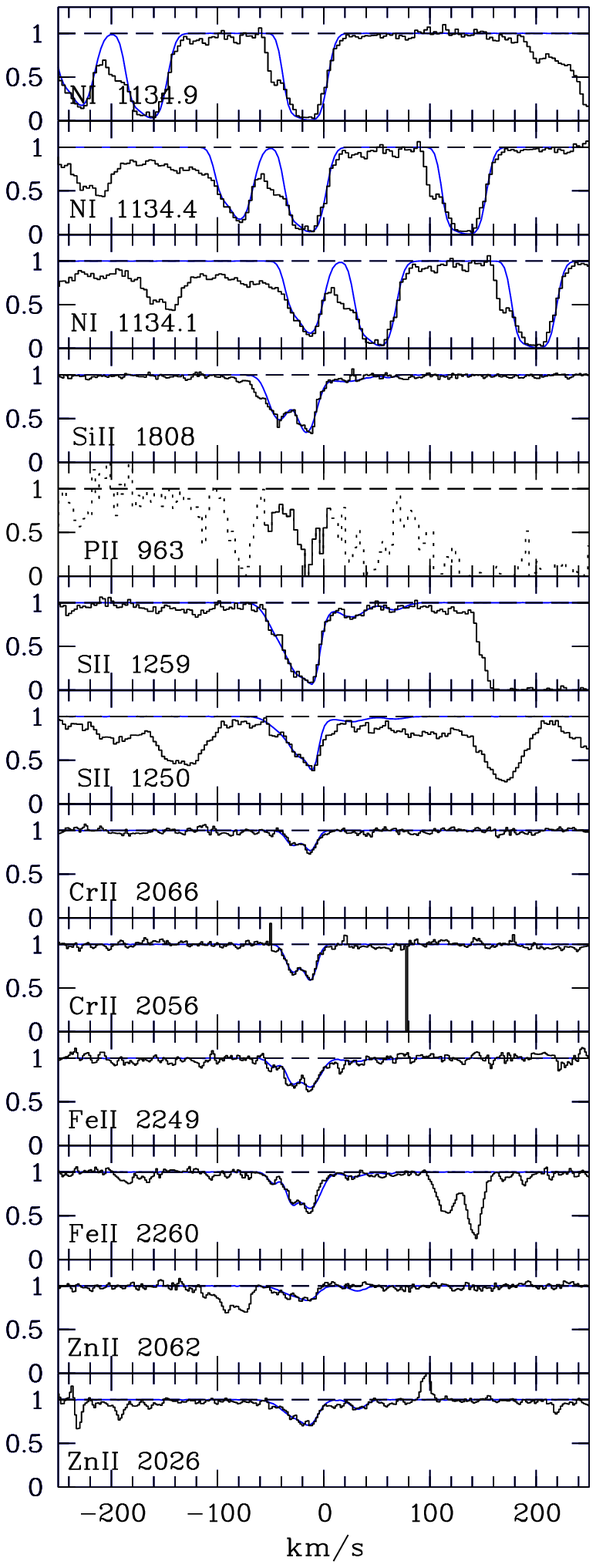} }}}
\caption{\label{fig_B} Same as in Fig.~\ref{fig_A} but for 
CTQ247B ($z_{abs}=2.5950$)}
\end{figure}

This DLA has a relatively simple velocity structure. Although the
strongest \ion{Si}{ii} and \ion{Fe}{ii} transitions have 5 velocity
components, these span only $\sim 80$ \kms. As expected for such a
high \ion{H}{i}, \ion{O}{i} is saturated and we consider it no further
here.  Conversely, \ion{N}{i} which is often barely detected exhibits
a large equivalent width in this system.  In fact, two of the
$\lambda$1134 triplet transitions are saturated.  However, the triplet
is relatively unblended, as are two of the \ion{S}{ii} triplet lines.
Fig.~\ref{fig_B} shows the nonsaturated transitions used and their
fits. There is an absorption feature at the expected
position of \ion{P}{ii} $\lambda 963$ but since we suspect
contamination by a \lya\ forest line we only provide an upper limit on 
$N(\ion{P}{ii})$. As for CTQ247A, we also provide an AODM
measurement for \ion{Mn}{ii}.

\paragraph{CTQ247C ($z_{abs}=2.6215$):}

\begin{figure}
\centerline{\rotatebox{0}{\resizebox{8.0cm}{!}
{\includegraphics{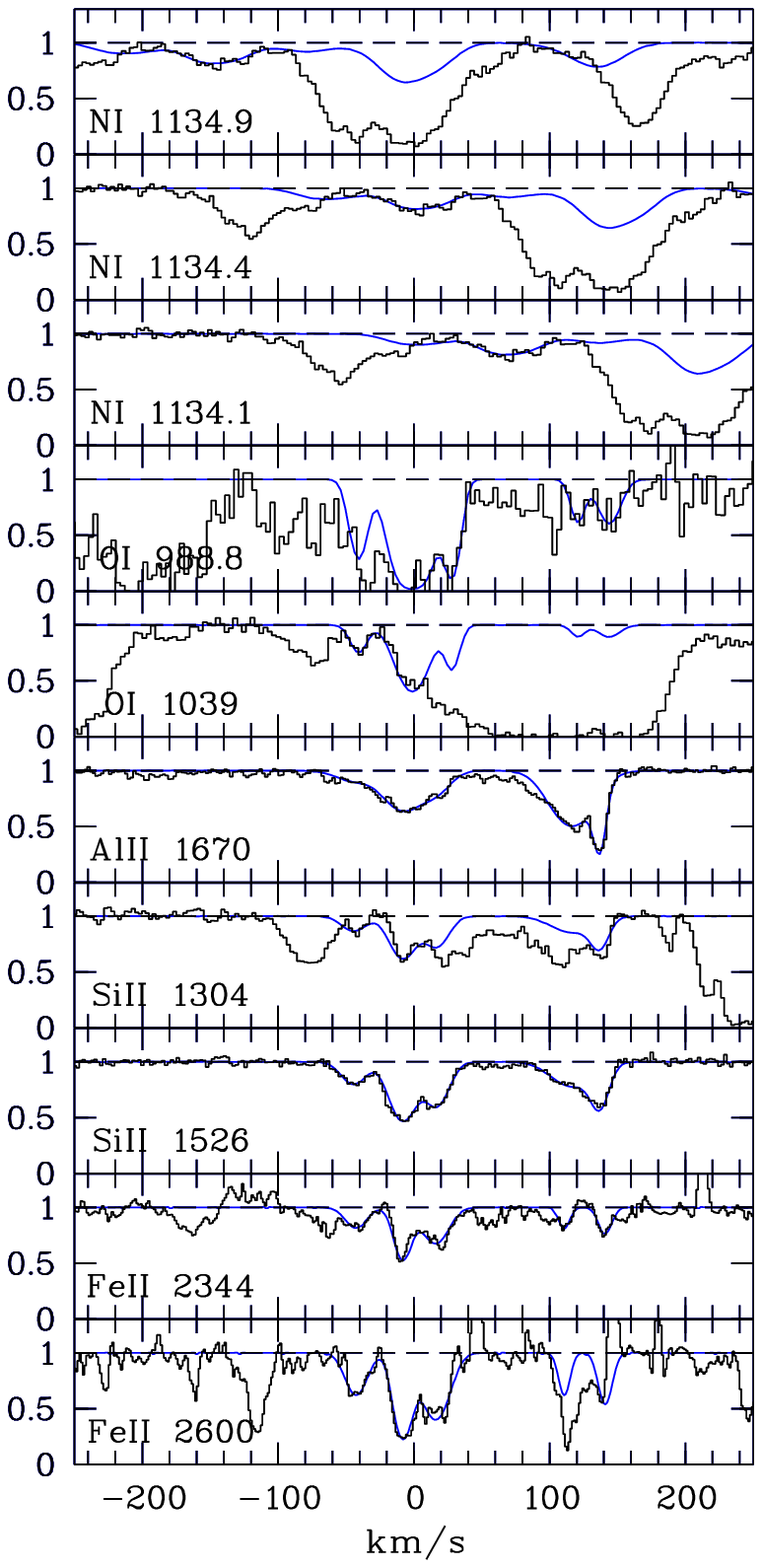} }}}
\caption{\label{fig_C} UVES spectrum showing nonsaturated transitions in 
CTQ247C ($z_{abs}=2.6215$)}
\end{figure}

This is the lowest \ion{H}{i} DLA of the triplet.  For this DLA the
bulk of the low ions are divided into 2 `sub-clumps' of 3 and 2
velocity components separated by $\sim 120$ \kms.  Fig.~\ref{fig_C}
shows the nonsaturated transitions used and their fits.  Since this
is the DLA with the lowest \ion{H}{i} of the 3, fewer species are
detected and limits are not very stringent.  For example, for sulphur
our $3\sigma$ detection limit is log $N$(\ion{S}{ii}) $<14.3$.  We have
attempted a fit for the \ion{N}{i} triplet, but a combination of many
weak components and Ly$\alpha$ blending makes this a very uncertain
result and we regard it as an upper limit due to possible contamination.  
On the other hand, the low
\ion{H}{i} allows us to profile fit 2 \ion{O}{i} transitions, for
which each one of the 5 components shows at least one nonsaturated
line. Also, we report one of the few existing measurements of
\ion{Al}{ii} in DLAs (line positions were held fixed at the
\ion{Si}{ii} values), and one AODM measurement for \ion{Al}{iii}.

\section{Abundances in MDLAs}
\label{abundances}

In Paper I we focused on the relative $\alpha$ to Fe peak abundances
in order to establish whether MDLAs may exhibit distinct SF histories.
Similarly, in Fig.~\ref{alpha} the solid circles depict [$\alpha$/Fe]
vs. [Fe/H] for the 3 MDLAs presented in this paper (CTQ247A, B, and
C), the double DLA Q2314-409A, B (Paper I), and Q2359$-$02 A, B 
(\cite{pw99}).  
The stars are DLAs taken from the literature.  Due to the potential
problem of blending between S lines and the \lya\ forest, we have compiled only
values obtained from echelle data.  In cases where
numbers are present in several references, we have preferred the
values by~\cite{ucsd1} for consistency.

In the following we show statistically  that the sub-class MDLA
is drawn from a different distribution to single DLAs. We then
discuss to which extent  dust and
ionization effects might produce such distinction, and what
possible systematic effects may be present in the 2 samples.

\subsection{Low $\alpha$/Fe in MDLAs}


\begin{table*}
\begin{center}
\begin{tabular}{ccccccc}\hline \hline\noalign{\smallskip}
  & \multicolumn{2}{c}{CTQ247A} & \multicolumn{2}{c}{CTQ247B} & \multicolumn{2}{c}{CTQ247C}  \\
X &  N(X) &  [X/H]&  N(X) &  [X/H] & N(X) &  [X/H]\\ \hline\noalign{\smallskip}
\ion{H}{i} & $21.13\pm0.10$ & ... & $21.09\pm0.10$  & ... & $20.47\pm0.10$ &...  \\
\ion{N}{i}  & $14.55\pm0.03$ & $-2.51\pm0.10$ & $15.07\pm0.02$ & $-1.95\pm0.10$ & $<14.36$ & $<-2.04$  \\ 
\ion{O}{i} &  ...& ... &...  & ... & $15.19\pm0.02$ & $-2.02\pm0.10$   \\
\ion{Al}{ii} & ...  & ... & ...  & ... & $12.88\pm0.07$ &$-1.97\pm0.12^b$    \\
\ion{Al}{iii} &  $12.81\pm0.05^a$& ... & $12.63\pm0.04^a$  & ... & $12.33\pm0.03^a$ & ...  \\
\ion{Si}{ii} & $15.32\pm0.04$ & $-1.37\pm0.11$ & $15.59\pm0.03$ & $-1.06\pm0.10$ & $13.99\pm0.06$ & $-2.04\pm0.12$  \\
\ion{P}{ii} & $<13.15$ & $<-1.54$ & $<12.98$ & $<-1.67$ & ... & ...  \\
\ion{S}{ii}  & $14.82\pm0.06$ & $-1.51\pm0.12$ & $15.19\pm0.05$ & $-1.10\pm0.11$ & $<14.34$ & $<-1.33$  \\
\ion{Cr}{ii} & $13.20\pm0.03$ & $-1.62\pm0.10$ & $13.37\pm0.04$ &$-1.41\pm0.11$&...&... \\
\ion{Mn}{ii} & $12.86\pm0.02^a$ & $-1.80\pm0.10$ & $12.87\pm0.01^a$ &$-1.75\pm0.10$ & ... & ...  \\
\ion{Fe}{ii} & $14.95\pm0.06$ & $-1.68\pm0.12$ & $15.15\pm0.02$ & $-1.44\pm0.10$ & $13.60\pm0.02$ & $-2.37\pm0.10$  \\
\ion{Fe}{iii} &  ...& ... & $<13.66$  & ... & ... & ...  \\
\ion{Ni}{ii} & $13.94\pm0.02$ & $-1.44\pm0.10$ & $13.86\pm0.31$ & $-1.48\pm0.33$ & ... & ...  \\
\ion{Zn}{ii} & $12.44\pm0.05$ & $-1.36\pm0.11$ & $12.68\pm0.02$ & $-1.08\pm0.10$ & ... & ...  \\ \hline
 & & & & & & \\
\end{tabular}
\caption{\label{abund_tab_ctq}Abundance measurements (and 3$\sigma$ upper limits)
for CTQ247. $^a$ uses AODM; $^b$ $N({\rm Al})=N(\ion{Al}{ii})+N(\ion{Al}{iii})$}
\end{center}
\end{table*}

%
%


\begin{figure}
\centerline{\rotatebox{0}{\resizebox{9.0cm}{!}
{\includegraphics{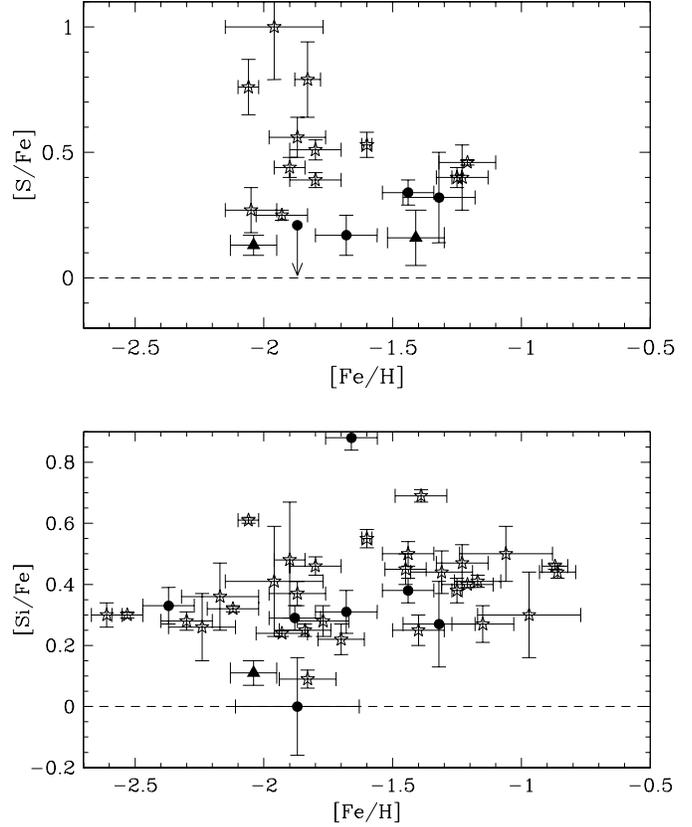} }}}
\caption{\label{alpha}
 Relative
ratios of $\alpha$/Fe for MDLAs (solid circles), 
the DLAs toward Q0201+1120 and Q0000$-$262 (solid triangles) and DLAs
taken from   the literature (stars).  All ratios have been corrected
to the following solar values: 
log (S/H)$_{\odot}=-4.80$, log (Si/H)$_{\odot}=-4.44$,
log (Fe/H)$_{\odot}=-4.50$ (Grevesse \& Sauval 1998). The data points
 shown  in this plot are listed in Table~\ref{tab_lit}.  } 
\end{figure}

\begin{table}
\begin{center}
      \[
         \begin{tabular}{lccccr}
            \hline
            \noalign{\smallskip}
{\rm QSO}&{\rm [Fe/H]}&{\rm [Zn/Fe]}&{\rm [Si/Fe]}&{\rm [S/Fe]}&{\rm Ref.}\\
            \noalign{\smallskip}
            \hline
            \noalign{\smallskip}
Q0013$-$004& -1.83  &   1.09  &    ...  &       0.79  &    m     \\  
Q0100+13   & -1.90  &   0.31  &    0.48 &       0.44  &    b,c  \\  
Q0149+33   & -1.77  &   0.10  &    0.28 &       ...   &    a,b     \\   
Q0201+3634 & -0.87  &   0.59  &    0.46 &       ...   &    a,n     \\ 
Q0216+0803A& -0.97  &   ...   &    0.30 &       ...   &    d	 \\ 
Q0216+0803B& -1.06  &   ...   &    0.50 &       ...   &    d	 \\ 
Q0255+00A  & -1.44  &   ...   &    0.50 &       ...   &    a       \\
Q0255+00B  & -2.05  &   ...   &    ...  &       0.27  &    a	   \\
Q0307$-$49 & -1.96  &   ...   &    0.41 &       1.00  &    e,h	   \\
Q0336$-$01 & -1.80  &   ...   &    ...  &       0.39  &    a	   \\
Q0347$-$38 & -1.80  &  $-0.10$&    0.46 &       0.51  &    a,h,i \\
Q0741+474  & -1.93  &   ...   &    0.24 &       0.25  &    a	   \\
Q0836+11   & -1.40  &   ...   &    0.25 &       ...   &    a       \\
Q0957+33A  & -1.45  &   ...   &    0.45 &       ...   &    a	   \\
Q0957+33B  & -1.87  &   ...   &    0.37 &       0.56  &    a	   \\
HE1104$-$18& -1.60  &   0.57  &    0.55 &       0.53  &    f	   \\
Q1108$-$07 & -2.12  &   ...   &    0.32 &       ...   &    a	   \\
Q1210+17   & -1.15  &   0.25  &    0.27 &       ...   &    a       \\
Q1215+33   & -1.70  &   0.41  &    0.22 &       ...   &    a	   \\
Q1223+17   & -1.84  &   0.22  &    0.25 &       ...   &    a	   \\
Q1331+17   & -2.06  &         &    0.61 &      0.76   &    a,b,o   \\
Q1409+095A & -2.30  &   ...   &    0.28 &       ...   &    k	   \\
Q1759+75   & -1.21  &   ...   &    0.40 &       0.46  &    a	   \\
Q1946+7658 & -2.53  &   ...   &    0.30 &       ...   &    a       \\
Q2206$-$19 & -0.86  &   0.45  &    0.44 &       ...   &    a	   \\
Q2206$-$19 & -2.61  &   ...   &    0.30 &       ...   &    a	   \\
Q2230+02   & -1.17  &   0.45  &    0.41 &       ...   &    a,b	   \\
Q2231$-$00 & -1.31  &   0.54  &    0.44 &       ...   &    b       \\
Q2237$-$060& -2.17  &   ...   &    0.36 &       ...   &    c,d	   \\
Q2243$-$603& -1.25  &   0.13  &    0.38 &       0.40  &    g	   \\
Q2343+1232 & -1.23  &   ...   &    0.47 &       0.40  &    c,l     \\
Q2344+12   & -1.83  &   ...   &    0.09 &       ...   &    a	   \\
Q2348$-$01A& -1.39  &   ...   &    0.69 &       ...   &    a	   \\
Q2348$-$01B& -2.24  &   ...   &    0.26 &       ...   &    a	   \\
\hline
CTQ247A    & -1.68  &  0.32   &    0.31 &       0.17  &    p    \\
CTQ247B    & -1.44  &  0.36   &    0.38 &       0.34  &    p    \\
CTQ247C    & -2.37  &  ...    &    0.33 &      $<$1.04&    p    \\
Q2314$-$409A&-1.32  &  ...    &    0.27 &       0.32  &  q   \\
Q2314$-$409B&-1.87  &  ...    &    0.00 &      $<$0.21   &  q   \\
Q2359$-$02A& -1.66  &  ...    &    0.88 &    ...      & b       \\
Q2359$-$02A& -1.88  &  ...    &    0.29 &    ...      & b       \\
            \noalign{\smallskip}
            \hline
         \end{tabular}
      \]
\caption{\label{tab_lit} Abundance ratios:  $^a$\cite{ucsd1};
  $^b$Prochaska \& Wolfe (1999);  
$^c$Lu, Sargent \& Barlow (1998); $^d$Lu et al. (1996);
  $^e$Dessauges-Zavadsky et al. (2001);
$^f$Lopez et al. (1999); $^g$Lopez et al. (2002); $^h$Bonifacio et
  al. (2001);
$^i$Levshakov et al. (2002);  $^j$Ge, Bechtold \& Kulkarni (2001);
  $^k$Pettini et al. (2002); 
$^l$Prochaska, Gawiser \& Wolfe (2001); $^m$Centurion et al. (2000);
  $^n$Prochaska \&   Wolfe (1996); $^o$Dessauges-Zavadsky, D'Odorico
  \& Prochaska (in
  prep.); $^p$This paper; $^q$Paper I
}  
\end{center}
\end{table}

As expected in DLAs, all points in Fig.~\ref{alpha} fall over the
[$\alpha$/Fe] $=0$ mark, as both S and Si are $\alpha$-chain
elements which are created mainly in massive stars on much shorter
scales than the long-lived iron producing stars.  The ratio
of [$\alpha$/Fe] is usually interpreted as an indicator of star
formation history and the rate at which the different elements
are released into the ISM. 
In addition to the MDLAs in Fig.~\ref{alpha}, we have also plotted
DLAs with other neighboring galaxies (usually detected via Lyman break
imaging) in the field.  These systems are Q0201+112 (\cite{q0201})
which has 4 identified galaxies in the vicinity, and Q0000-262  
(\cite{Molaro}) for which \cite{ste96} found 2
Lyman break galaxies within $\Delta z=0.05$ of the absorber.  

The remarkable property to note in Fig.~\ref{alpha} is that all MDLAs
systematically show low [S/Fe] values with a small scatter ($<0.2$ dex).  To
quantify this impression, we have performed a modified KS test that
provides the likelihood of MDLAs with [S/Fe] measurements to be
drawn from a different distribution to single DLAs.  We created 1000
realizations of a simulation which recreates the [S/Fe] distribution
of literature
DLAs and MDLAs including the 1 $\sigma$ quoted errors.  For
each DLA, an error was drawn at random from a Gaussian distribution
and added to the observed value\footnote{We treat the upper limit
toward DLA B in Q2314$-$409 as a detection; a lower value than this
does not affect the KS probability}.  A KS test was then run on each
of the 1000 datasets and the likelihood that the KS probability
$P_{\rm KS}$ was greater than a certain confidence level was calculated
from the number of realizations.  In this fashion, we overcome 2
problems associated with a normal KS statistics: the small number of
MDLAs we have at our disposal and the exclusion of measurement errors.

According to our simulations, the likelihood $p$ that the two samples are
drawn from different distributions is $p=1.00$ for $P_{\rm KS}>95$ \% (i.e.
every one of the 1000 realizations gives a $P_{\rm KS}>95$ \%),
$p=1.00$ for $P_{\rm KS}>98$ \%, and
even as high as $0.97$ for $P_{\rm KS}=99$ \%.  This indicates that
the consistently  
low [S/Fe] values in MDLAs and other multiple systems are
statistically significant.  Since Si and S are both produced in
oxygen burning reactions, [S/Si] $=0$ in all observed Galactic disk
stars (\cite{Chen}) and we may expect to see similar patterns between
S and Si in DLAs.  However, the MDLA [Si/Fe] values have a slightly
larger dispersion and there are a number of low literature [Si/Fe]
values, which renders the two populations less distinguishable.  This
may be explained because low $N(\ion{S}{ii})$ lines are easily \
contaminated in the Ly$\alpha$ forest, whereas low $N(\ion{Si}{ii})$
can be much more easily measured.    
If we consider only
the literature DLAs with Si $and$ S measurements, the KS probability
that the two [Si/Fe] samples are consistent is 11\%, as opposed to
32\% if all are included.  There is an
obvious outlier in the [Si/Fe] distribution associated with system
Q2359$-$02A; in the following section we argue that this system has
anomalously large dust content compared with the rest of the MDLAs and
literature DLAs.  Discounting this exceptionally dust depleted
system reduces the KS probability even further of all [Si/Fe]
measurements to 9\%.  Therefore, although there is a hint that [Si/Fe]
may be low in MDLAs, the evidence is not as convincing as that for S.
We explore this further in the following section.

\subsection{How dust and/or ionization might affect the observed
ratios} 

\begin{figure}
\centerline{\rotatebox{270}{\resizebox{5.0cm}{8.0cm}
{\includegraphics{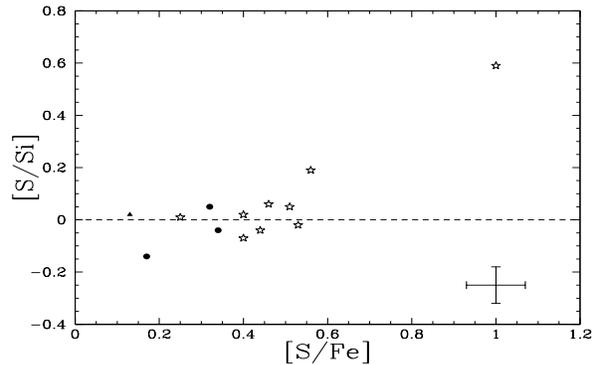} }}}
\caption{\label{fig_dust}
[S/Si] vs. [S/Fe] in MDLAs (circles), single DLAs (stars), and in the
DLA toward Q0000$-$262 (triangle). 
}
\end{figure}

Before discussing the implications of the observed low [$\alpha$/Fe]
ratios in terms of a nucleosynthetic origin, we must first rule out
other (non-intrinsic) systematics.  In Paper I, we have already
discussed the issue of dust and photoionization in the double DLA
toward Q2314-409.  Here, we embark upon a similar assessment for the
CTQ247 triplet.  In particular, we investigate whether systematics may
cause (a) the $\alpha$/Fe ratio to be lower in MDLAs than in single
DLAs; and (b) the $\alpha$ under-abundance in MDLAs to be more evident
in S rather than Si.

First, we consider dust depletion.  Although disentangling dust
depletion from pure nucleosynthesic effects in DLAs is still matter of
debate, some broad trends can be established: (1) In no instance is S
known to be depleted in the Galactic ISM (e.g. Savage \& Sembach
1996).  Conversely, Si is mildly depleted in most environments and Fe
is very easily incorporated into dust.  Therefore, dust depletion will
cause an increase in both [S/Fe] and [Si/Fe], the effect being most
pronounced in the former ratio, so [S/Si] $>0$. (2) Zn is undepleted;
thus, assuming the same nucleosynthetic origin as Fe, one expects
[Zn/Fe] $>0$ in dusty systems.

Both in CTQ247A and B the
[Zn/Fe] shows a significant departure from solar, indicating that dust
\textit{is} present.  This is also true for Q2314$-$409, the other
MDLA for which a Zn abundance is determined; in all cases [Zn/Fe] $>$
0.3, typical of other DLAs.  Therefore, any downward correction due to
dust would further emphasize the low $\alpha$/Fe in these MDLAs.  It
is noteworthy that Q2359$-$02A, which has the largest [Si/Fe] of all,
also shows an unusually large dust content, [Zn/Fe]=+0.88, explaining
its anomalously large [Si/Fe].  Not only is Q2359$-$02A relatively
highly depleted compared to other MDLAs, but also in comparison with
other DLAs which mostly have [Zn/Fe] $<$ 0.6. 

Can low dust content in MDLAs {\it compared} with single DLAs be
responsible for lower [S,Si/Fe] in the former?  Fig.~\ref{fig_dust}
shows [S/Si] vs. [S/Fe] for the (few) systems where both elements are
available, including 3 MDLAs. It can be seen that apart from one DLA with
an extremely large [S/Fe], there is no strong trend
for systems with lower [S/Si] (supposedly less dusty) to show lower
[S/Fe], suggesting that low S/Fe ratios are not strongly effected by dust. 
Indeed, the large S abundance measured in Q0307$-$49 might be due to
\lya\ blending (\cite{bonifacio01}) and  must  be considered with
caution.  Therefore, low S/Fe is not obviously correlated with low dust
content, at least not to the accuracy we can presently measure this
ratio. In conclusion, it does not seem from the present sample that a
general trend of low $\alpha$/Fe values be driven by any effect of
dust, nor do MDLAs have atypical dust content.

\begin{figure}
\centerline{\rotatebox{0}{\resizebox{9.0cm}{!}
{\includegraphics{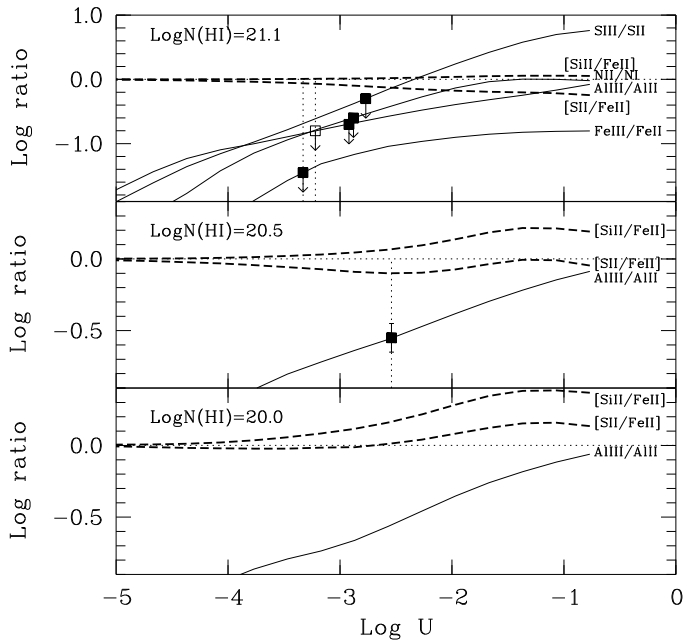} }}}
\caption{\label{cloudy}
Ratio of doubly to singly ionized species as a function of ionization
parameter $U$ from CLOUDY models with a Haardt \&
Madau (\cite{Haardt}) ionizing spectrum.
Top panel: Model with $\log N(\ion{H}{i})=21.1$ and gas metallicity 
$\log Z=\log Z_{\sun} - 1.5$ compared with data  points  for CTQ247A (open
square) and CTQ247B (filled squares).  Middle panel:
The measured \ion{Al}{iii}/\ion{Al}{ii} ratio for CTQ247C, and a model
using $\log N(\ion{H}{i})=20.5$ and metallicity $\log 
Z=\log Z_{\sun} - 2.5$.  Bottom panel:
A model  using $\log N(\ion{H}{i})=20.0$, the column density cutoff in
our definition of MDLAs, and
metallicity $\log Z=\log Z_{\sun} - 1.5$. In all panels the dashed
lines indicate solar-corrected ratios while the solid lines are for
column density ratios.  }
\end{figure}

We now turn to ionization.  Given the large $N(\ion{H}{i})$ of CTQ247A
and B, it is highly unlikely that ionization corrections are required
(\cite{Viegas}).  Nonetheless, Prochaska et al (2002c) found
significant corrections in a DLA with $\log N(\ion{H}{i}) = 20.8$, so
there are occasional cases where photoionization plays a part even in
relatively large column density systems.  In order to constrain the
ionization parameter for CTQ247A and B we use CLOUDY (\cite{Ferland})
models with a $\log N(\ion{H}{i}) = 21.1$, a Haardt-Madau ionizing
spectrum (\cite{Haardt}) at $z=2.3$ and a metallicity of $-1.5$ solar,
and use the observed ratios to constrain the ionization parameter.
For CTQ247A, $R \equiv \log N(\ion{Al}{iii})/N(\ion{Al}{ii})$ provides
the most stringent limit on log $U$, whereas the ratio of
\ion{Fe}{iii}/\ion{Fe}{ii} provides most information on the ionization
of CTQ247B.  In both cases, the limiting ionization parameter is found
to be $\log U < -3.2$, see upper panel of Fig.~\ref{cloudy}.
Consequently, this limits the correction to [S/Fe] to be less than
0.07 dex. Note that a correction this extent is in agreement
with the subsolar [S/Si] observed for the 3 MDLAs in
Fig.~\ref{fig_dust} and  brings this ratio into line with solar,
within the measurement errors.

CTQ247C has a significantly lower \ion{H}{i} column density.  Our
spectrum does cover the \ion{Fe}{iii} $\lambda 1122$ transition but
the limit is not very meaningful. However, we can compare \ion{Al}{ii}
and \ion{Al}{iii} whose ratio has been shown to exhibit a steady trend
with $N(\ion{H}{i})$, a higher fraction of \ion{Al}{iii} being present
for lower $N(\ion{H}{i})$ (\cite{Vladilo1}).  We find $R = -0.55$
which is typical of moderate column density DLAs, $\log N(\ion{H}{i})
> 20.5$.  Repeating the above CLOUDY models with the appropriate
\ion{H}{i} confirms that the corrections to [S/Fe] will not exceed
$0.1$ dex in this MDLAs (bottom panel in
Fig.~\ref{cloudy}). Therefore, we would need to make a small upward
correction to [S/Fe] of up to about 0.1 dex but this does not alter
our conclusions.

We note that the MDLAs toward Q2359$-$02 have both high $R\approx
-0.3$, indicating relatively high ionization. According to our CLOUDY
models, however, the corrections for [S/Fe] are still negligible
for the corresponding ionization parameter, 
while [Si/Fe] requires downward corrections of $0.1$--$0.2$ dex.

From a more general point of view, Fig.~\ref{cloudy} tells us that for
a typical $\log N(\ion{H}{i})=20.5$ DLA the corrections are downward
for Si/Fe and upward for S/Fe, whereas at the low end of \ion{H}{i} column
densities the [Si/Fe] ratio is significantly more sensitive to
ionization than  [S/Fe] and both corrections are downward
(cf.~\cite{ucsd3}).  Ionization will therefore require downward
corrections in [Si/S] but these will not necessarily contribute to the 
wider spread of MDLA [Si/Fe] values we observe in Fig.~\ref{alpha}.

\begin{figure}
\centerline{\rotatebox{0}{\resizebox{7.0cm}{!}
{\includegraphics{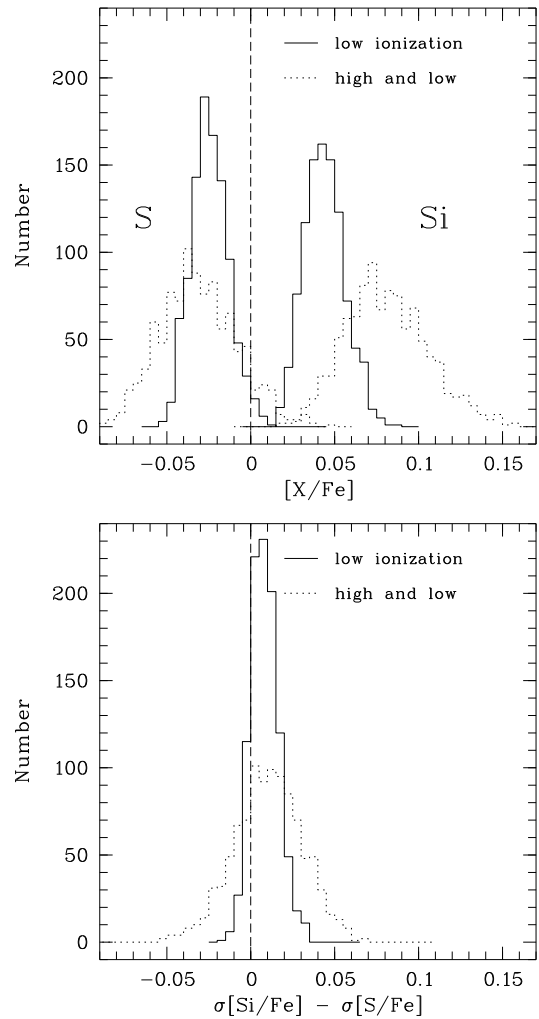} }}}
\caption{\label{mc}Simulated [S/Fe] and [Si/Fe] distributions (top panel) and
difference between their scatters (bottom panel). 
The solid (dotted) histogram shows DLA simulations drawn from a sample of DLAs
with $-4<\log U<-3$ ($-4<\log U<-2$ )}
\end{figure}

We can quantify the effect of ionization on the 
comparative dispersion between [Si/Fe] and [S/Fe] by performing
Monte-Carlo simulations with a grid of
CLOUDY models. To this end, we constructed 2 samples of $N=500$ CLOUDY
models each, with $-4<\log U<-3$ (`low-ionization' sample) and $-4<\log
U <-2$ (`high+low-ionization' sample), and \ion{H}{i} column density and
metallicity values uniformly distributed in the ranges
$19.4<\log N(\ion{H}{i})<22.0$, $-2.4<\log(Z/Z_{\sun})<-0.8$
\footnote{Of course, uniformly distributed values of \ion{H}{i} and $Z$ 
do not represent the observed distribution of DLAs, this range is
purely for the purpose of investigating the effects of ionization.}. 
To simulate our
observations we selected randomly $1\,000$ sets of 20 DLAs each from the
two samples. Their  [X/Fe] distributions are shown in the upper panel
of Fig.~\ref{mc}. The {\it difference} between the scatters in
[Si/Fe] and [S/Fe] was calculated as  $\Delta S\equiv \sigma{\rm [Si/Fe]} -
\sigma{\rm [S/Fe]}$,  where $\sigma{\rm [X/Fe]}$ is the standard
deviation of 
each dataset. The bottom panel of the figure shows the
$\Delta S$ distribution for both samples. For the
low+high-ionization sample the probability $P(\Delta S>0)$ of having a wider
scatter in  [Si/Fe] is roughly $0.7$, while $P(\Delta S>0)= 0.85$
if the DLAs are drawn exclusively from the low-ionization
sample. This is an indication that ionization does affect the
distributions differentially, although the differences are generally  
below measurement errors.


\subsection{Other possible systematic effects} 

Finally, other possible systematic effects we have to beware of are:
(1) Column densities: our
column densities  come from $\chi^2$ fits whereas the majority of 
literature DLAs have AODM 
measurements. However, it is well established that both approaches
give essentially same results provided the lines are nonsaturated as
is our case. (2) Solar abundances: solar abundances have undergone
some significant revisions in recent years; all the points in figures
~\ref{alpha} and~\ref{nitrogen} have been put  on a common solar
scale. (3) Atomic data: some $f$-values have changed in recent
years, most
notably \ion{Fe}{ii} $\lambda 1611$. However, this should not  be an
issue as long as very few systems rely on only one transition \footnote{e.g.,
Q2206$-$19A is the only DLA in the sample that relies solely on
\ion{Fe}{ii} $\lambda 1611$}. (4) 
Blending:  \ion{S}{ii} transitions are normally in the forest and  therefore
particularly    susceptible to contamination.  This is especially a
worry if the AODM is 
used since it  would cause larger $N({\rm S})$ if contamination is
present. Since  40\% of literature S measurements 
comes from the UCSD database which uses the AODM,  in theory we could have
compared our data with overestimations of S/Fe. However, checking in
\cite{ucsd1} on a case-by-case basis shows that this is unlikely to be
a problem because the 
possibility of contamination was carefully assessed in each case.  Even
excluding [S/Fe] points determined by AODM from Fig.~\ref{alpha} still
shows the low S trend in MDLAs. (5) Instrument: almost all literature
abundances come from HIRES data whereas we have used UVES at a
slightly higher resolution. However, some high $\alpha$/Fe UVES values in
Table~\ref{tab_lit} and the low $\alpha$/Fe HIRES value in the MDLAs
toward Q2359$-$02 show that a possible bias due to different data type
is unfounded. 

In summary, we have found that MDLAs exhibit significantly lower
[S/Fe], and to a less striking extent, [Si/Fe] ratios than single DLAs 
taken from the literature.  We have argued that these low values are not
driven by ionization differences or dust depletion, although an observational
bias may be partially responsible for the clearer abundance distinction in S
compared to Si.  Having excluded the main systematic effects, these
results therefore imply a nucleosynthetic origin for differences in
abundance ratios between MDLAs and single DLAs.

\section{Discussion}
\label{conclusions}

\subsection{The Nature of MDLAs}

If the typically low values of [$\alpha$/Fe] of MDLAs are truly
nucleosynthetic in origin, this may indicate that these systems either
represent a particular galaxy type or that they share a common
evolutionary or environmental link.  The first issue to address is 
whether the `multiplicity' is indeed due to more than one distinct galaxy.
\cite{Schaye} has suggested that some DLAs may be associated with
winds, in which case MDLAs may be caused by outflowing material
from the parent DLA.  Indeed, \cite{nbf98} have shown that large
scale galactic winds can produce column densities equal to those of DLAs.
This could be tested by studying abundances
in systems that are effectively transverse multiple systems, that
is DLAs observed in more than one line of sight separated by many tens
of kpc.  The closest example we have of this in the current dataset
is the case of Q0201+1120 (\cite{q0201}) which is located
in a concentration with at least four other galaxies and indeed
shows low [S/Fe], possible evidence that the abundance trend is
symptomatic of a group environment rather than simply
line of sight velocities within a single object.

However, this evidence remains vague
in our context of MDLAs as these 
imaging studies deal only with  DLAs without
line-of-sight companions. Deep imaging and subsequent spectroscopy of
the fields toward CTQ247 and Q2314-409 
as well as control fields around QSOs harboring single
DLAs is therefore required in order  to pursue a 
comparison of possibly different environments. 
At higher redshifts, finding evidence for
cluster environments of single DLAs (and even the DLA absorbers
themselves) has proven difficult (\cite{Gawiser}; \cite{ProchaskaAJ}),
so much work is still to be done in this direction. 
Similarly, although several DLA imaging studies have been carried out 
(e.g., \cite{LeBrun}), we lack studies of the clustering properties 
in the DLA fields at $z < 1.0$ and there are currently no known MDLAs
at low redshift. 

\subsection{Truncated star formation in MDLAs?}

As suggested in Paper I, the low $\alpha$/Fe abundances observed in MDLAs
might be due to suppressed SF in the galaxies harboring
the absorbing gas.  We now raise the
question as to whether environment may cause these observed
abundance anomalies and speculate upon the possibility that MDLAs
arise in `protogalactic groups'.

There is a clear trend of galaxies in rich environments out to $z \la 0.5$ to 
exhibit suppressed SF, not just in dense cluster cores, but also in clusters
out to 2-3 virial radii (\cite{bal97}; Lewis et al. 2002) and perhaps
out to several Mpc (Pimbblet et al 2001).   The
line diagnostics of these galaxies indicate that radial trends are
consistent with an age sequence in which the last episode of star
formation happened most recently in the outermost galaxies (\cite{bal99}).  However, the physical mechanisms governing this effect are not
clear; violent processes such as ram pressure stripping (e.g., 
\cite{Couch} and references therein) and passive
exhaustion of gas supply (e.g., \cite{bal99}) are the 
two main possibilities.  The recent finding by Lewis et al. (2002) that
local density is the dominant parameter in suppressing SF
rates indicates that extreme cluster-scale processes such as 
ram-stripping by the ICM play a minor role.  These results also point 
to the possibility that reduction in SF need not take place within 
the gravitational bounds of existing virialized
clusters and may occur already in
looser groups before they are accreted onto larger structures.
This ties in with cluster histories at lower redshifts, because
although \cite{dres99} find a significant
post-starburst population in clusters, there is evidence that this
fraction is not greater than among the field population (\cite{bal99}).

If the same physical processes are on-going at high redshifts, then
suppressed SF may be evident in galaxies associated with
early proto-clusters, such as those identified by Steidel et al (1998),
even though canonical, virialized clusters have yet to form.
Although the abundance ratios for MDLAs that we have discussed here
are distinct from `field' DLAs, they have typical metallicities
for their redshift (c.f. Pettini et al 1999) and moderate dust content.
These properties are indications that some chemical maturity has
indeed taken place, and MDLAs are similar to other DLAs in terms of
their general enrichment.  However, the low [$\alpha$/Fe] is an indication
that what sets MDLAs apart is that they have been evolving quiescently,
without any further major SF episodes, over the last $\sim$
500 Myrs.

\subsection{Other evidence for quiescent star formation}

\begin{figure}
\centerline{\rotatebox{0}{\resizebox{8.0cm}{!}
{\includegraphics{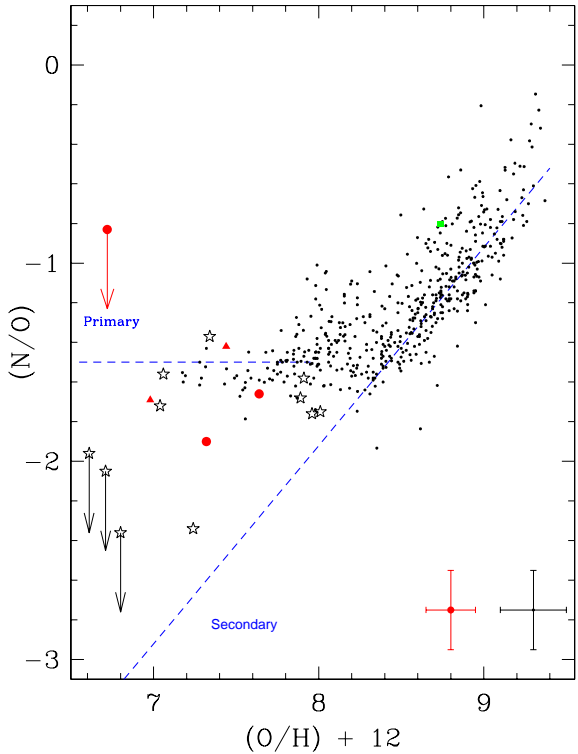} }}}
\caption{\label{nitrogen}
The N/$\alpha$ ratio in MDLAs (circles), 'field' DLAs (stars)
and \ion{H}{ii} regions (dots; see Pettini et al. 2002 for 
references). The triangles are for the 2 DLAs with companion LBGs.
}
\end{figure}

If the low $\alpha$/Fe ratios observed in MDLAs are to be explained by
quiescent SF, then we should expect to see its signature in other
elemental ratios that also act as `cosmic clocks'.  The most promising
possibility is the N/$\alpha$ ratio which traces the primary and secondary
contributions from stars of different masses (\cite{hek00}).
Fig.~\ref{nitrogen} shows the N/$\alpha$ ratio in MDLAs, `field' DLAs
and \ion{H}{ii} regions (Pettini et al. 2002).  
We only used measurements of O or S in the
plot and corrected S values by the solar S/O ratio (1.54 dex). 
Prochaska et al. (2002b) have suggested a bimodal distribution
of  N/$\alpha$ and although this needs to be confirmed by more
data, it is clear that a number of DLAs exhibit low ratios.  For
MDLAs, the ratios are all high, with no points (N/O) $< -1.9$, supporting the
quiescent SF scenario. It perhaps indicates more chemical maturity
than average, but also that
the role of massive stars  in polluting the ISM of MDLAs may be
less important.    


The other established trend with metallicity in Galactic stars is the
odd-even effect. Although this is not technically a chronometer in
the same sense as N and the alpha elements, there is a correlation with 
metallicity which may be due to  metallicity dependent yields (\cite{gp00}).
At our disposal we have [Mn/Fe] $=-0.12\pm0.08$ (CTQ247A) and
[Mn/Fe] $=-0.31\pm0.03$ (CTQ247B), the former of which is among the 
highest values in
DLAs of this metallicity (\cite{lbp02}; Dessauges-Zavadsky et al
2002).  If these 
high ratios are not due to dust depletion (Fe is more depleted in the 
local ISM than
Mn) or ionization, they show a mild odd-even effect.  Analogously,
[Al/Si] $=0.07\pm0.12$ in CTQ247C, the highest value ever measured,
and [Al/Si] $=-0.28\pm0.18$ in B2314$-$409B (Paper I), which is among
the highest values after the $z=2.154$ MDLA 
toward Q2359$-$02 (\cite{ucsd2}). 
Altogether, the odd-even effect seems to be mild in those
MDLAs where we can measure it. 
Since models of massive star yields
give good agreement with observations of {Mn/Fe} in Galactic stars
(\cite{gp00}; and references therein) and also DLAs (\cite{lbp02}),
the lack of  
odd-even effect at low metallicity in MDLAs indicates again that
high-mass stars have not been the dominant  metal pollutant in
MDLAs. This may be interpreted as 
further evidence that there have not been several epochs of SF
building on the enriched interstellar media from previous episodes;
further support of suppressed SF.

\section{Summary}

We have studied a sample of 7 DLAs with line-of-sight companions,
`multiple' DLAs in our nomenclature, and 2 DLAs arising in transverse
groups. We have shown that the relative abundances in these two
sub-samples are unusual as compared with single DLAs. In particular,
the [S/Fe] and [Si/Fe] ratios are statistically lower than literature
DLAs, although the effect is more striking in [S/Fe].  We suggest that
this difference is due to (1) an observational bias against measuring
low S column densities because weak \ion{S}{ii} are usually lost in
the forest; and (2), to a lesser degree, the larger scatter induced in
[Si/Fe] by photoionization.  Besides low $\alpha$/Fe ratios, MDLAs
also exhibit a mild odd-even effect and relatively high N/$\alpha$
ratios. We interpret these results as truncated star formation in
MDLAs.  If the multiplicity of these DLAs is due to grouping in
physical space, a thesis we cautiously support, environment may be the
cause of the quiescent SF scenario, just as is observed in more nearby
clusters and groups of galaxies.

\begin{acknowledgements}

We are grateful to the anonymous referee for many qualified comments
and suggestions on a first version of this paper. We also thank Eric
Gawiser for  fruitful discussions and comments,  
Michael Rauch for allowing us to use the FORS data of CTQ247, Max
Pettini for providing us with the template for Fig.~\ref{nitrogen},
Mirka  Dessauges-Zavadsky for communicating abundances in advance of
publication, 
and Jason Prochaska, whose database at {\tt
http://kingpin.ucsd.edu/\~{}hiresdla/} we have consulted. SL
acknowledges support from the Chilean {\sl Centro de Astrof\'\i sica}
FONDAP No. 15010003, and from FONDECYT grant N$^{\rm o} 3\,000\,001$.

\end{acknowledgements}


\begin{thebibliography}{}


\bibitem[Ballester et al. 2000]{bal00}
	Ballester, P., Modigliani, A., Boitquin, O., Cristiani, S.,
	Hanuschik, R., Kaufer, A., Wolf, S., 2000, ESO Messenger,
	101, 31

\bibitem[Balogh et al. 1999]{bal99}
        Balogh, M., Morris, S., Yee, H. K. C., Carlberg, R. G.,
	Ellingson, E., 1999, ApJ, 527,54

\bibitem[Balogh et al. 1997]{bal97}
        Balogh, M., Morris, S., Yee, H. K. C., Carlberg, R. G.,
	Ellingson, E., 1997, ApJL, 488, L75

\bibitem[Bonifacio et al. 2001]{bonifacio01}
Bonifacio, P.,  Caffau, E.,  Centurion, M.,  Molaro, P. \&   Vladilo,
G., 2001, MNRAS, 325, 767


\bibitem[Centurion et al. 2000]{Centurion}
Centurion, M.,  Bonifacio, P.,  Molaro, P. \&  Vladilo, G. 2000, ApJ,
536, 540 

\bibitem[Chen et al. 2002]{Chen}
Chen, Y. Q., Nissen, P. E., Zhao, G. \&  Asplund, M., 2002, A\&A, 390,
225

\bibitem[Chengalur \& Kanekar 2000]{Chengalur}
      Chengalur, J. N. \& Kanekar, N., 2000, MNRAS, 318, 303

\bibitem[Couch et al. 2001]{Couch}
        Couch, W., Balogh, M., Bower, R., Smail, I., Glazebrook, K.,
\& Taylor,  M., 2001, ApJ., 549, 820


\bibitem[Dessauges-Zavadsky, Prochaska \& D'Odorico (2002)]{dpo02}
        Dessauges-Zavadsky, M., Prochaska, J., \& D'Odorico, S., 2002,
	A\&A, 391, 801


\bibitem[Dessauges-Zavadsky et al. 2001]{dz01}
Dessauges-Zavadsky, M., D'Odorico, S., McMahon, R. G., Molaro, P.,
Ledoux, C.,  Peroux, C. \& Storrie-Lombardi, L. J, 2001, A\&A, 370, 426


\bibitem[Dressler et al. (1999)]{dres99}
        Dressler, A., Smail, I., Poggianti, B., Butcher, H.,
	Couch, W., Ellis, R., Oemler, A., 1999, ApJS, 122, 51

\bibitem[Ellison \& Lopez (2001)]{el01}
        Ellison, S. L., \& Lopez, S., 2001, A\&A, 380, 117, Paper I


\bibitem[Ellison et al. 2001a]{q0201}
        Ellison, S. L.,  Pettini M., Steidel C. C., Shapley A., 2001a, ApJ,
	549, 770

\bibitem[Ellison et al. 2001b]{corals1}
	Ellison, S. L., Yan, L., Hook, I., Pettini, M., Wall, J., Shaver, P.,
	2001b, A\&A, 379, 393

\bibitem[Evrard et al. 2002]{Evrard02} Evrard, A. E., MacFarland,
  T. J., Couchman, H. M. P., Colberg, J. M., Yoshida, N., White,
  S. D. M., Jenkins, A., Frenk, C. S., Pearce, F. R., Peacock,
  J. A. \& Thomas, P. A., 2002, ApJ, 573, 7 

    \bibitem[Ferland 1993]{Ferland}
      Ferland, G. J. 1993, University of Kentucky, Physics Department
      Internal Report

   \bibitem[Gawiser et al. 2001]{Gawiser} 
Gawiser, E., Wolfe, A. M., Prochaska, J. X., Lanzetta, K. M., Yahata,
 N., \&  Quirrenbach, A. 2001, ApJ, 562, 628


\bibitem[Goswami \& Prantzos 2000]{gp00}
        Goswami, A., \& Prantzos, N., 2000, A\&A, 359, 191

\bibitem[Grevesse \& Sauval (1998)]{gs98}
        Grevesse, N., \& Sauval, A.J. 1998, Space Sci Rev, 85,
        161

   \bibitem[Haardt \& Madau 1996]{Haardt} 
      Haardt, F., \& Madau, P. 1996, ApJ, 461, 20


\bibitem[Haines, Campusano \& Clowes 2003]{hcc03}
        Haines, C., Campusano, L. E. \& Clowes, R., 2002,
        astro-ph/0301473 

\bibitem[Henry, Edmunds and Koeppen 2000]{hek00}
	Henry, R. B. C., Edmunds, M. G., \& K\"{o}ppen, J.
	2000, ApJ, 541, 660

\bibitem[Holweger (2001)]{hol01}
        Holweger, H., 2001, in Solar and Galactic Composition, ed. R.
	Wimmer-Schweingruber, (Berlin: Springer), 23

\bibitem[Hou, Boissier \& Prantzos 2001]{Hou}
        Hou, J. L., Boissier, S. \& Prantzos, N. 2001, A\&A, 370, 23  



   \bibitem[Le Brun et al. 1997]{LeBrun} 
    Le Brun, V., Bergeron, J., Boisse, P. \& Deharveng, J. M., 1997,
    A\&A, 321, 733

\bibitem[Ledoux, Bergeron \& Petitjean 2002]{lbp02}
	Ledoux, C., Bergeron J., \& Petitjean, P., 2002,
	A\&A, 385, 802

\bibitem[Levshakov et al. 2002]{lev02}
Levshakov, S. A., Dessauges-Zavadsky, M.,  D'Odorico, S. \& Molaro,
P., 2002, ApJ, 565, 696

   \bibitem[Lewis et al. 2002]{Lewis} 
      Lewis I., Balogh, M., De Propris, R., et al. 2002, MNRAS, 334,
      673 

     \bibitem[Lopez et al. 1999]{lopez99} 
      Lopez S., Reimers D., Rauch M., Sargent
      W. L. W. \& Smette A., 1999, ApJ 513, 598


     \bibitem[Lopez et al. 2001]{Lopez01}     
       Lopez, S., Maza, J., Masegosa, J., \&  Marquez, I., 2001, A\&A 366,
       387  

     \bibitem[Lopez et al. 2002]{lopez02}     
       Lopez, S., Reimers, D. D'Odorico, S. \& Prochaska, J. X.,
       2002, A\&A 385, 778  

\bibitem[Lu, Sargent and Barlow 1998]{lsb98}
	Lu, L., Sargent, W. L. W., Barlow, T.A.
	1998, AJ, 115, 55

    \bibitem[Lu et al. 1996]{Lu96}
      Lu, L., Sargent, W. L. W., Barlow, T. A., Churchill, C. W.,
      Vogt, S. 1996, ApJS 107, 475


    \bibitem[Molaro et al. 2001]{Molaro}
      Molaro, P., Levshakov, S. A., D'Odorico, S., Bonifacio, P., \&
      Centurion, M.,  2001,  ApJ 549, 90 
 
\bibitem[Nulsen, Barcons \& Fabian (1998)]{nbf98}
        Nulsen, P. E. J., Barcons, X., Fabian, A. C., 1998, MNRAS, 301, 168

    \bibitem[Peroux et al. 2001]{Peroux}
      Peroux, C. Storrie-Lombardi, L. J., McMahon, R. G., Irwin, M. \&
      Hook, I. M.,   2001, AJ, 121, 1799


\bibitem[Pettini et al. 2002]{petinni02}
Pettini, M., Ellison, S. L., Bergeron, J. \& Petitjean, P., 2002, A\&A,
391, 21 

\bibitem[Pettini et al. 1999]{pet99}
	Pettini, M., Ellison, S. L., Steidel, C. C., Bowen, D. V.
	1999, ApJ, 510, 576


\bibitem[Pimbblet et al (2001)]{pim01}
        Pimbblet, K, Smail, I., Edge, A., Couch, W., O' Hely, E.,
	Zabludoff, A., 2001, MNRAS, 327, 588

\bibitem[Prochaska et al. 2002a]{ProchaskaAJ}
   Prochaska, J X., Gawiser, E., Wolfe, A. M., Quirrenbach, A.,
Lanzetta, K. M.,  Chen, H.-W, Cooke, J., \& Yahata, N. 2002a, AJ, 123,
2206 

\bibitem[Prochaska et al (2002b)]{no02}
        Prochaska, J. X., Henry, R., O'Meara, J., Tytler, D., Wolfe, A.,
	Kirkman, D., Lubin, D., Suzuki, N., 2002b, ApJ, accepted

\bibitem[Prochaska et al. 2002c]{ucsd3}
        Prochaska, J. X., Howk, J. C., O'Meara, J. M., Tytler, D.,
Wolfe, A., Kirkman, D., Lubin, D. \& Suzuki, N.  2002c, ApJ, 571, 693  

\bibitem[Prochaska \& Wolfe 2002]{ucsd2}
        Prochaska, J. X., \& Wolfe, A., 2002, ApJ, 566, 68


\bibitem[Prochaska et al. (2001)]{ucsd1}
Prochaska, J. X., Wolfe, A. M., Tytler, D., Burles, S.,  Cooke, J.,
 Gawiser, E., Kirkman, D.,  O'Meara, J. M., \& Storrie-Lombardi,
 L. ApJS, 2001, 137, 21 

\bibitem[Prochaska, Gawiser \& Wolfe 2001]{prochaska01}
Prochaska, J. X., Gawiser, E. \& Wolfe, A. M., 2001, ApJ, 552, 99  


\bibitem[Prochaska \& Wolfe 1999]{pw99}
	Prochaska, J.X., \& Wolfe, A.M.
	1999, ApJS, 121, 369

\bibitem[Prochaska \& Wolfe 1996]{pw96}
	Prochaska, J.X., \& Wolfe, A.M.
	1996, ApJ, 470, 403

    \bibitem[Quashnock, vanden Berk \& York 1996]{Quashnock96} 
     Quashnock, J. M., vanden Berk, D. E. \&  York, D. G., 1996, ApJ
     472, 69 

\bibitem[Rigopoulou et al.\ 2002]{rig02}
	Rigopoulou, D., Franceschini, A., Aussel, H., Genzel, R.,
	Thatte, H., Cesarsky, C., 2002, ApJ, 580, 789

    \bibitem[Savage \& Sembach (1991)]{Savage1} 
      Savage, B. D. \&{} Sembach, K. R., 1991, ApJ 379, 245

\bibitem[Savage and Sembach 1996]{ss96}
	Savage, B. D. \& Sembach, K. R., 1996, ARA\&A, 34, 279

\bibitem[Schaye (2001)]{Schaye}
        Schaye, J., 2001, ApJL, 559, L1

\bibitem[Steidel et al (1998)]{ccs98}
        Steidel, C.C., Adelberger, K.L., Dickinson, M., Giavalisco, M.,
        Pettini, M. \& Kellogg, M. 1998, ApJ, 492, 428.

\bibitem[Steidel et al. (1996)]{ste96}
         Steidel, C., Giavalisco, M., Pettini, M., Dickinson, M., Adelburger,
	 K., 1996, ApJL, 462, L17

    \bibitem[Viegas 1995]{Viegas}
      Viegas, S. M., 1995 MNRAS, 276, 268

   \bibitem[Vladilo et al. 2001]{Vladilo1}
      Vladilo, G., Centuri\'on, M., Bonifacio, P., \&  Howk,
      J. C., 2001,  ApJ 557, 1007

   \bibitem[Vladilo 2002]{Vladilo02}
      Vladilo, G., 2002, ApJ, 569, 295

   \bibitem[Williger et al. 2002]{Williger02}
       Williger, G. M., Campusano, L. E., Clowes, R. G. \&  Graham,
       M. J., 2002, ApJ, 578, 708

\bibitem[Wolfe et al. 1986]{Wolfe86}
        Wolfe, A. M., Turnshek, D. A., Smith, H. E. \& Cohen,
R. D. 1986, ApJS 61, 249 


\bibitem[Zaritsky et al. (1997)]{zar97}
        Zaritsky, D., Smith, R., Frenk, C., White, S. D. M., 1997
	ApJ, 478, 39

\end{thebibliography}
\end{document}